%% file: ms_2.tex
\documentstyle[citedal,psfig]{mn2e}

\newif\ifAMStwofonts

\title{Mass Modelling with Minimum Kinematic Information}
\author[Dalia Chakrabarty]
       {Dalia Chakrabarty$^{1}$ \thanks{E-mail:dalia.chakrabarty@nottingham.ac.uk}\\
$^{1}$ School of Physics $\&$ Astronomy, University of Nottingham, Nottingham NG7 2RD, U.K.}

\date{\today}

\begin{document}

\maketitle

\label{firstpage}

\begin{abstract}
\input{abstract.tex}
\end{abstract}


\section{Introduction}
\label{sec:intro}
\noindent
The recovery of the total mass distribution in early type systems is a
knotty problem. The obvious reason for this is lack of information;
photometry offers access to the stellar mass but the deduction of the
arrangement of the total mass in a system remains elusive, unless
kinematic information is invoked. Even when velocity information is
available, the implementation of the same is besotted with problems of
varying degrees of subtleness (\cite{merritt93}, \cite{merritt96},
\cite{dalpras}, \cite{gerhardgenzel}).

Of the many debacles that plague the processing of kinematics, the
worst is perhaps the mass anisotropy degeneracy; even in the
simplistic case of a spherical system, the mass distribution can be
known only as a function of the anisotropy in velocity space
(\cite{bible}). Getting an independent handle on anisotropy is
difficult, if not impossible in most cases. \scite{cote87} attempt
this for the system of globular clusters in M87, by constraining the
total mass from the work of \scite{mclaughlin99}. Their conclusion
indicates that the velocity distribution of this system in M87 is
``almost perfectly isotropic''. This is consistent with the appraisal
of isotropy in globular cluster systems in other galaxies
(\cite{zepf00}, \cite{cote03}).

An attempt is also underway to produce a fully anisotropic algorithm,
that is similar in flavour to the existing code CHASSIS that works by
implementing kinematic data of tracers to recover the potential and
distribution function from which the tracer sample is most likely to
have been drawn (\cite{dalpras}, \cite{dalzwart}). However, a large
number of kinematic observations is required by this proposed
extension. Nonetheless, what is heartening is that such ambitious data
sets are less of a figment of the astronomer's reverie today, owing to
the huge kinematic surveys that are planned or already in operation.

Another technique that is often employed in mass modelling of
elliptical galaxies involves the utilisation of X-ray
measurements. The procedure suffers from uncertainties in the observed
temperature profile and simplification in terms of the assumption of a
single component in the X-ray emitting gas and of hydrostatic
equilibrium. A comprehensive review of various the shortcomings of the
conventional mass modelling techniques is presented in
\scite{mamon05}.

Thus, it is clear that at the present moment, utilisation of kinematic
information in mass modelling, suffers from egregious drawbacks. Even
more crucially, such data is hard to come by in many external
galaxies, including those at higher redshifts. For this latter class,
the implementation of velocity information in formalisms that stem
from a stellar distribution function, is all the more debatable since
systems outside the local universe are most probably not relaxed.

This indicates an opening for a technique that is able to extricate
the mass profile of early type systems from photometry, using as
little velocity information as possible. Moreover, the used kinematic
information should not be used in a procedure that relies on the
rigorous employment of the equilibrium stellar distribution function.
Such a scheme is advanced in this short report. 

It merits mention that any such mass modelling formalism for
elliptical galaxies will bear the potential of shedding light on the
connection between the luminous and dark matter content in such
systems. In the past, such correlations have been explored in spiral
galaxies (\cite{sancisi04}, \cite{verheijen01} and
\cite{noordermeer06}), but very little is known about the distribution
of the total mass in ellipticals; \scite{padmanabhan04} and
\scite{lintott06} discuss sample galaxies in this context. The scheme
discussed here will indicate the same in model systems.

The paper is arranged as follows. The next section provides an
exposure to the details of the proposed scheme.  In
Section~\ref{sec:sersic}, the Sersic model considered in the paper is
introduced. Section~\ref{sec:results} is devoted to the results obtained
with the toy galaxy models, while the following section
(Section~\ref{sec:allowed}) delineates the galaxy models that can be
tackled with the proposed scheme from those that are outside the
purview of this scheme. Section~\ref{sec:m87} discusses the details
of the results obtained with the data for M87, as obtained from the
ACS Virgo Cluster Survey (ACSVCS), (\cite{laura06}), and comparison of
the recovered mass model with those predicted by \scite{cote87} and
\scite{romkoch87}, (from the analysis of kinematic data of tracer
populations in M87).  The last section is devoted to the discussions
of some of the salient points that were raised in the other sections
and a summary of the results.

\section{Details of the Formalism}
\label{sec:scheme}
\noindent
We work under the assumption that the surface brightness profile of a
galaxy is known to us. The central velocity dispersion ($\sigma_0$) is
also considered known from observations.

\subsection{Deprojection}
\noindent
The first step is to deproject the observed brightness profile into the
three dimensional luminosity density distribution. This can be done
with the non-parametric deprojection algorithm DOPING, which is the
acronym for Deprojection of Observed Photometry using an INverse
Gambit (\cite{doping06}).

This algorithm can perform the deprojection under a general geometry
that needs to be specified as an input. The other fundamental input is
the inclination. The measurements that DOPING processes are the
surface brightness measurements and the projected ellipticity profile,
at different values of the major axis coordinate $x$, i.e. the surface
brightness map of the galaxy. This is used to provide the three
dimensional luminosity density distribution, the line-of-sight
projections of which are the best reproductions of the brightness
data, along any azimuth on the plane of the sky. The deprojection of
the observed photometry for systems with varying intrinsic shape
requires the implementation of a regularisation technique, though
systems in which the ellipticity is uniform are more easily dealt
with. It needs to be mentioned that DOPING can in principle, deal with
changes of eccentricity and twist angle, with galactocentric distance.

The errors on the deprojected luminosity profile are the errors of the
analysis, which measures the $\pm$1-$\sigma$ extent of the excursion of the
algorithm around the global maxima in the likelihood function;
likelihood is maximised on spotting the luminosity density
distribution that projects best into the data. The observational
errors can be also incorporated into the analysis.

\subsection{Models}
\noindent
In this paper, we attempt to mass model a set of toy elliptical
galaxies. In terms of measurable quantities, these model systems are
ascribed a brightness profile $I(x)$ and a projected ellipticity
$\epsilon$, that we choose to fix at 0.4, (implying that for edge-on
viewing, the intrinsic eccentricity $e$=0.8). All the toy galaxies are
assumed to have been observed in the SDSS $I$-band and placed at the
distance of Virgo (considered as 17Mpc). 

Multiple runs are carried out with the toy galaxies being assigned
varying values of central velocity dispersion ($\sigma_0$), from which
the local mass-to-light ratio at the centre of the galaxy is
calculated in a way discussed below. These observables are the inputs
into the mass modelling system that is proposed in the paper. The
recovered total mass density distribution is then compared to the true
mass density profiles ($\rho_t(x)$) of these models. The degree of
overlap between the predicted and known mass distributions is then
visually tracked, in order to confirm the success and limitations of
the proposed mass modelling scheme.

The models are described by a luminous component (luminosity density
$L(x)$) lying embedded in a dark halo that is chosen to be NFW type
(density: $\rho_{NFW}(x)$.  Thus, the true total mass density of a
model galaxy can be written as:
\begin{equation}
\rho_t(x) = \displaystyle{
                          \rho_{NFW}(x) +
                          \alpha10^{\frac{L_\odot - L(x)}{2.5}}
                         },
\label{eqn:total}
\end{equation}
where the dark matter density is 
\begin{equation}
\rho_{NFW}(x) = \displaystyle{\frac{M_s}{4\pi{x}(x+r_s)^2}}
\label{eqn:nfw}
\end{equation}
and $\alpha$ is a measure of the luminous matter fraction, such that
its reciprocal indicates the halo fraction in the toy galaxy. Also,
$L_\odot$ is the absolute magnitude of the Sun in the same waveband as
the brightness observations.

We treat two sets of model brightness distributions: the cored and the
Sersic type profiles. As far as the cored brightness distributions are
concerned, the following prescription is used. The brightness profile
of the test galaxies is extracted from the line-of-sight integration
of an analytically chosen luminosity density function. (Thus, the
deprojection of this brightness data can be compared to the known
luminosity density to confirm the correctness of the
deprojection). This analytical luminosity density distribution
$L_t(x)$ is chosen to have the form:
\begin{equation}
L_t(x) = \displaystyle{\frac{A}
                        {\left[r_c^2 + x^2 + {y^2}/({1-e^2})+z^2\right]^{1.5}}
                      },
\label{eqn:L}
\end{equation}
where $A$ is the central luminosity density and $r_c$ the core radius
of the test galaxy in question. Thus, various toy surface brightness
data sets are prepared, with varying amplitude and core radii. The
deprojection is carried out under the assumptions of oblateness and an
edge-on viewing angle.

\begin{figure}
\centerline{
\psfig{figure=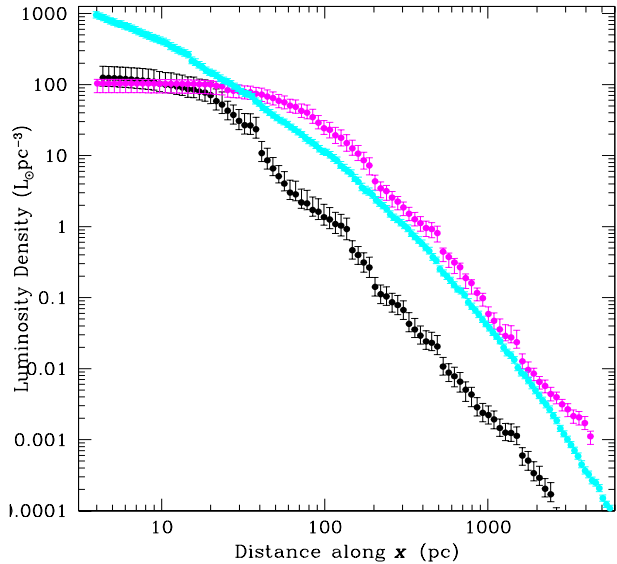,height=8.0truecm,width=8.5truecm}}
\caption{\small{Deprojected luminosity density profiles of toy
galaxies with core radii of about 7.1pc (in black) and 22.4pc (in
magenta), recovered by DOPING from model brightness profiles that were
constructed from the analytical luminosity distribution given in
Equation~\ref{eqn:L}, normalised to 100 L$_\odot$pc$^{-3}$. The
Sersic luminosity density is shown in cyan, normalised to 1000
L$_\odot$pc$^{-3}$.}}
\label{fig:cored_den}
\end{figure}

We normalise the deprojected luminosity density profiles of all the
model galaxies such that the central value of $L(x)$ is about 100
L$_\odot$pc$^{-3}$, leaving the models distinguishable from each other
in terms of the core radius only.  Each such model luminosity density
profile is then ported to the mass modelling formalism (discussed
below) and the outcome is compared to the known mass density
distribution of this toy galaxy. Figure~\ref{fig:cored_den} shows the
normalised luminosity distributions of the cored galaxies (of core
radius 7.07pc and about 22.4pc) that have been used in our
experiments.

For the Sersic models, the brightness is chosen to be described by a
Sersic index of 5.3 and a scale length of about 108pc. This central
density is normalised to 1000 L$_\odot$pc$^{-3}$, i.e. 1 order of
magnitude higher than the normalisation factor for the cored galaxies.

\subsection{Raw Two-stepped Discontinuous $M/L$ Profile}
\label{sec:ml}
\noindent
Once the deprojection is performed, the next stage is to choose a
total $local$ mass-to-light ratio ($M/L$) profile that scales the
obtained luminosity density profile in such a way that it would
hopefully emulate the total mass density. Or at least, such is the
aim, as in other procedures that hope to connect photometry to a mass
model. We suggest selection of a discontinuous, two-step local
$M/L$:
\begin{eqnarray}
M/L &=& \Upsilon_{in} \quad {\rm if} x \leq x_{in}\\  \nonumber
M/L &=& \Upsilon_{out} \quad {\rm if} x > x_{in}
\label{eqn:ml}
\end{eqnarray}
The luminosity density is scaled by this mass-to-light ratio profile
and then $smoothed$. Of course, the success of the scheme depends on
the selection of $x_{in}$, $\Upsilon_{in}$ and
$\Upsilon_{out}$. 

Additionally, we need to be cautious that this method is susceptible
to spurious answers at distances where the dark matter halo dominates
the system. Thus, we aim to produce a mass profile over a radial range
that is not too extensive (not exceed about 3$X_e$, where $X_e$ is a
measure of the effective radius of the galaxy). We ascertain $X_e$ by
measuring the slope ($m$) of the straight line that is fit to the plot
of core-removed log$_{10}$(I) against $x^{1/4}$: $X_e =(-3.33/m)^4$. 


So, what would amount to a judicious prescription for the selection of
the free parameters in this formalism? If we were to assume that mass
follows light in elliptical galaxies, then we could set
$\Upsilon_{in}=\Upsilon_{out}$ and $x_{in}$ could be set to any value in
the range of [0, 3$X_e$]. But there is no reason to believe that mass
does follow light in elliptical galaxies. In fact, gas kinematics and
X-ray studies indicate otherwise; \scite{buote} report that the
hypothesis that mass follows light can be rejected at 96$\%$ for their
oblate models for NGC720 and 98$\%$ for their prolate
models. \scite{bertola} point out that the run of the mass-to-light
ratio with galactocentric radius in ellipticals is similar to that in
spirals.

Kinematics too rule out the possibility of a constant $M/L$ at 95$\%$
confidence (\cite{gerhardkronawitter}). More recently, the analysis of
the kinematics of planetary nebulae observed in five elliptical
galaxies, by the Planetary Nebula Spectrograph, has led to the
announcement that these galaxies are drastically short of dark matter
in the outer parts (\cite{pnsscience},
\cite{romanowsky04}). \scite{napolitano05} advance a parameter that
measures the logarithmic slope of the mass-to-light ratio in a sample
of early type galaxies, to find that it increases with the stellar
mass. They conclude that the distribution of dark matter or its
fraction changes with radius.

\subsubsection{Jump Radius}
\noindent
Thus, we are left to make as good a choice for the parameters as
possible, from whatever can be learnt about the system from the
minimum of observables. Now, there is no observational constraint that
can suggest a value for the jump radius $x_{in}$, at which the above
two-step local $M/L$ profile should suffer its discontinuity. However,
one obvious length scale in the problem is $X_e$. So we can choose to
impose a certain multiple of $X_e$ as the jump radius; in fact,
experiments indicate that a good value of $x_{in}$ is 3$X_e$. This is
not a particularly bad choice either, given that by 3 effective radii,
the dark matter content can be expected to begin to exert its
influence. Having fixed $x_{in}$, we now proceed to use whatever
observable information there may be available to constrain
$\Upsilon_{in}$ and $\Upsilon_{out}$.

\subsubsection{Inner $M/L$}
\noindent
One measure that is often available from observations is the central
velocity dispersion in galaxies. $\sigma_0$ can be utilised to get a
handle on the local mass-to-light ratio in the inner parts of the
galaxy (i,e, $\Upsilon_{in}$). This can be done by implementing
$\sigma_0$ that is measured at $x=x_0$ (say), in the virial
theorem. However, the estimation of the mass enclosed within $x_0$
($M_0$), from which the average mass density ($\rho_0$) inside $x_0$
can be calculated in this way, will be mired in errors owing to a lack
of information about the degree of anisotropy present at $x=x_0$, as
well as the extent of deviation from sphericity at $x_0$. Virial
theorem suggests that under the assumption of velocity isotropy,
\begin{equation}
\displaystyle{3\sigma_0^2} = \displaystyle{\frac{GM_0}{\frac{1}{2}x_0}}
\label{eqn:virial}
\end{equation}
and oblateness and edge-on inclination imply
\begin{equation}
M_0 = \displaystyle{\frac{4\pi}{3}\rho_0{x_0^3\sqrt{1-e^2}}}
\end{equation}
This estimated $\rho_0$ will deviate from the true inner mass density,
as explained above. Anisotropy will spuriously lower the estimated
enclosed mass and therefore the calculated $\rho_0$, for a given
$\sigma_0$. For example, \scite{padmanabhan04} suggest that the
dynamical mass enclosed within the projected half-light radius ($R_e$)
of the massive ellipticals in their SDSS sample is given as 1.65$^{2}$
times $\sigma_0^2{R_e}/G$, with a systematic error of 30$\%$
(\cite{lintott06}).

\begin{figure*}
\centerline{
\psfig{figure=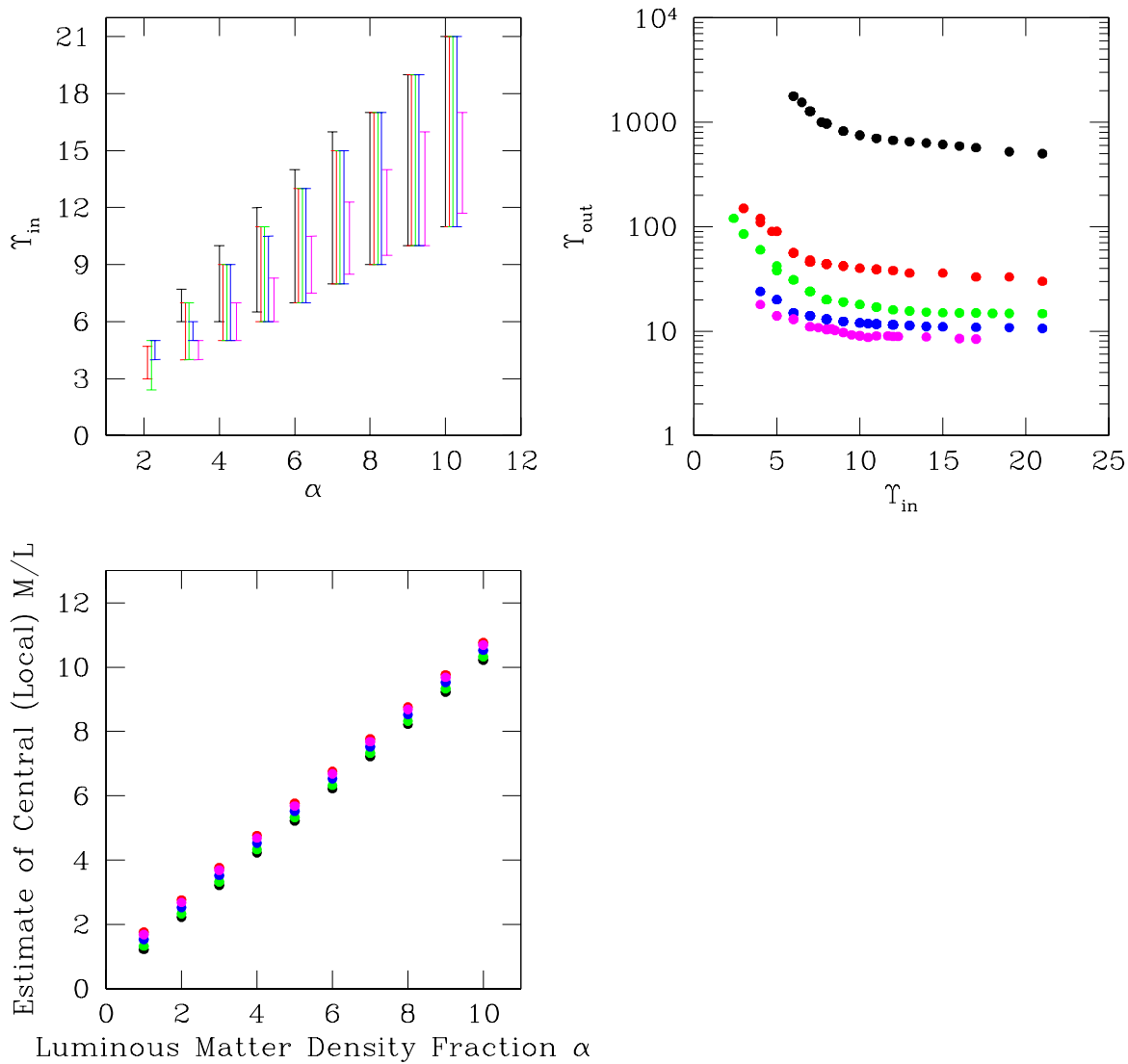,height=16.0truecm,width=16.5truecm}}
\caption{\small{Lower left panel: the luminosity matter fraction
$\alpha$, plotted as a function of the estimated central
$M/L$. Results for $r_c$=7.1pc are shown in black, for 14.4pc in red,
for 17.3pc in green, for 20pc in blue and for 22.4pc in magenta. Top
left panel: figure showing the allowed ranges in $\Upsilon_{in}$, as a
function of $\alpha$ (which measures the estimated central local
$M/L$), for cored galaxies of varying core radii; the colour indexing
used here is the same as above. Top right panel: $\Upsilon_{out}$
plotted as a function of $\Upsilon_{in}$.}}
\label{fig:alpha_gamin_gamout}
\end{figure*}

Expectedly, the central $M/L$ ($M/L(0)$) has a one-to-one and
monotonically increasing relationship with the luminous matter
fraction index $\alpha$ (lower left panel of
Figure~\ref{fig:alpha_gamin_gamout}). As shown in the figure, $M/L(0)$
is well represented by $\alpha$, though the deviation of
$[M/L(0)]/\alpha$ from unity exists - this ratio decreases from about
0.22 to 0.07 as we increase $r_c$ from 7.1pc to 22.4pc. {\it We will
use this result to justify our choice of $\alpha$ for parametrising
the estimated central $M/L$.}

We scan through a range of estimated central $M/L$ values
(i.e. $\alpha$) and monitor the success of the proposed technique; in
the process, the ranges of $\Upsilon_{in}$ and $\Upsilon_{out}$
corresponding to a given choice of the central $M/L$ value are
recorded.

Now, experiments with our model galaxies of a given core radius $r_c$
indicate that for a given $\alpha$, there exists a range of values for
$\Upsilon_{in}$ that will result in compatibility of the estimated and
known mass density profiles (upper left panel in
Figure~\ref{fig:alpha_gamin_gamout}).  This band of allowed
$\Upsilon_{in}$ values for a given $\alpha$, varies with core radius of
the normalised luminosity density profile in hand. For our cored
galaxies, this band of $\Upsilon_{in}$ values that are allowed at a
given $\alpha$, range from just above the corresponding value of
$\alpha$ to approximately 2 times this $\alpha$, with this factor
varying only weakly with core radius $r_c$.

If we assume the value of 2.72 between the true value of the mass
enclosed within an effective radius and $\sigma_0^2{R_e}/G$, as
suggested by \scite{padmanabhan04} (and used by \cite{lintott06}),
then the true central $M/L$ should be about 1.82 times the virial
estimate. Thus, the range of 2$\alpha$ that is allowed in
$\Upsilon_{in}$, for a given estimate of $M/L(0)$, appears to be
inclusive of real ellipticals, at least the systems that are as
massive as those in the sample of \scite{padmanabhan04}.

If it so happens that a certain galaxy is far too anisotropic at the
centre, such that the true $M/L(0)$ exceeds the upper bound on the
$\Upsilon_{in}$ allowed at the $\alpha$ at hand, then the mass density
profile that will result from using the highest value of the permitted
$\Upsilon_{in}$ in our mass modelling formalism, will not concur with
the profile that results from using the lowest allowed
$\Upsilon_{in}$. Thus, in such circumstances, when our scheme fails,
we will be aware that it has done so.

Thus, it may be summarised that even though the virial theorem is
invoked to help constrain $\Upsilon_{in}$, there is enough scope
available within the formalism to accommodate for the problems caused
by velocity anisotropy in traditional kinematic modelling. This and
other flexibilities offered by the proposed formalism are discussed in
greater detail in Section~\ref{sec:discussions}.

\subsubsection{Outer $M/L$}
\noindent
Our experiments indicate that it is possible to achieve a one-to-one
correlation between $\Upsilon_{in}$ and $\Upsilon_{out}$, for a toy
configuration of a given core radius. In other words, irrespective of
$\alpha$, whenever a value of $\Upsilon_{in}$ is chosen,
$\Upsilon_{out}$ is fixed. This relationship between $\Upsilon_{in}$
and $\Upsilon_{in}$ is depicted on the top right panel in
Figure~\ref{fig:alpha_gamin_gamout}. 

Actually, there exists a range of $\Upsilon_{out}$ for a given
$\Upsilon_{in}$, with the extent of this range maximising itself at
values of $\Upsilon_{in}$ that lie in the middle of the range that is
allowed for a given $\alpha$.  However, to keep matters simple, we
present only the value of $\Upsilon_{out}$ that lies at an extrema of
this range, for a given $\Upsilon_{in}$.

Analytical fits to the plots suggest that $\Upsilon_{out}$ varies as
$A_1\exp[-\Upsilon_{in}/\tau] + A_0$, where $A_1$ and $A_0$ are
constants that are determined from fitting this exponentially decaying
function to the plot of $\Upsilon_{out}$ against $\Upsilon_{in}$;
$\tau$ measures the decay rate. Fitting this functional form to the
experimentally obtained plots, suggest power-law dependences of both
the additive and multiplicative constants, on $r_c$:
\begin{enumerate}
\item $A_1\sim{r_c}^{-5.31}$, 
\item $A_0\sim{r_c}^{-3.75}$ and 
\item $\tau$ has a weak dependence on $r_c$, according to
$\tau=1.005^{-6}r_c^4 +1.85$.
\end{enumerate}

That the inner and outer local $M/L$ ratios are connected, betrays a
relationship between the luminous and dark matter content in these toy
elliptical galaxies. This facet of the models and the conclusions
that we can draw on the basis of these results for elliptical galaxies
in general, are further explored in Section~\ref{sec:discussions}.

\subsubsection{Smoothing}
\label{sec:smooth}
\noindent
Once the local $M/L$ is specified, it is used to scale the luminosity
density distribution, in order to obtain a mass density distribution,
which is indeed rendered discontinuous by the process of its
formulation. Thus, the next stage is to smooth this mass density
distribution. 

We choose to smooth this density profile with a triangle filter of
filter size given by 2$X_e$, i.e. two successive applications of a box
filter of filter size corresponding to the length $X_e$. It is
appropriate that the smoothing filter size be a function of one of the
characteristic scale-lengths of the system; $X_e$ is the most obvious
candidate that satisfies this criterion. Thus, the trends in
$\Upsilon_{out}$ with variation in $\Upsilon_{in}$, for the different
normalised toy galaxy configurations, have been extracted from
experiments performed with a smoothing filter size of $X_e$. However,
other smoothing windows are also usable - except that the exact nature
of the relationship between $\Upsilon_{out}$ and $\Upsilon_{out}$ will
then be different from what is prescribed above, for a given $r_c$.

\section{Sersic Galaxies}
\label{sec:sersic}
\noindent
All the toy models that are discussed above correspond to examples of
surface brightness profiles that are cored. In order to investigate
the full applicability of our mass modelling technique, a Sersic type
galaxy is now examined.

We consider a toy Sersic surface brightness distribution, described by
a central density of 1000 L$_{\odot}$pc$^{-3}$, a scale length of
about 100pc and a Sersic index of 5.3. When deprojected under the
assumption of sphericity, a luminosity density profile along the
semi-major axis is recovered. The central density chosen for this toy
model is 1000 L$_{\odot}$pc$^{-3}$ instead of 100 L$_{\odot}$pc$^{-3}$
(as in the cored models), since the proposed mass modelling technique
does not work for Sersic models at the lower normalisation.

Once normalised in this way, a range of central (local) mass-to-light
ratio values are imposed, in order to scan the range of parameters
over which our scheme is viable. As before, the central $M/L$ values
are imposed in terms of the parameter $\alpha$ (the luminous matter
fraction). The raw two-step local $M/L$ profile that is imposed upon
this luminosity density is again chosen to be described by an inner
$M/L$ of $\Upsilon_{in}$ and an outer $M/L$ of $\Upsilon_{out}$, with
the jump radius given by 3$X_e$. 

In these Sersic models, $X_e$ is defined as the slope ($m$) of the
straight line that is fit to the plot of log$_{10}$(I) against
$x^{1/4}$: $X_e = (-3.33/m)^4$. For the toy model at hand, we find
$X_e\approx$726.8pc.

Once the two-stepped local $M/L$ profile is fully characterised, the
deprojected luminosity density is scaled by it and the resulting mass
density distribution is smoothed. As before, it is smoothed by two
successive applications of a boxcar smoothing filter of filter size
given by $X_e$.

\begin{figure*}
\centerline{
\psfig{figure=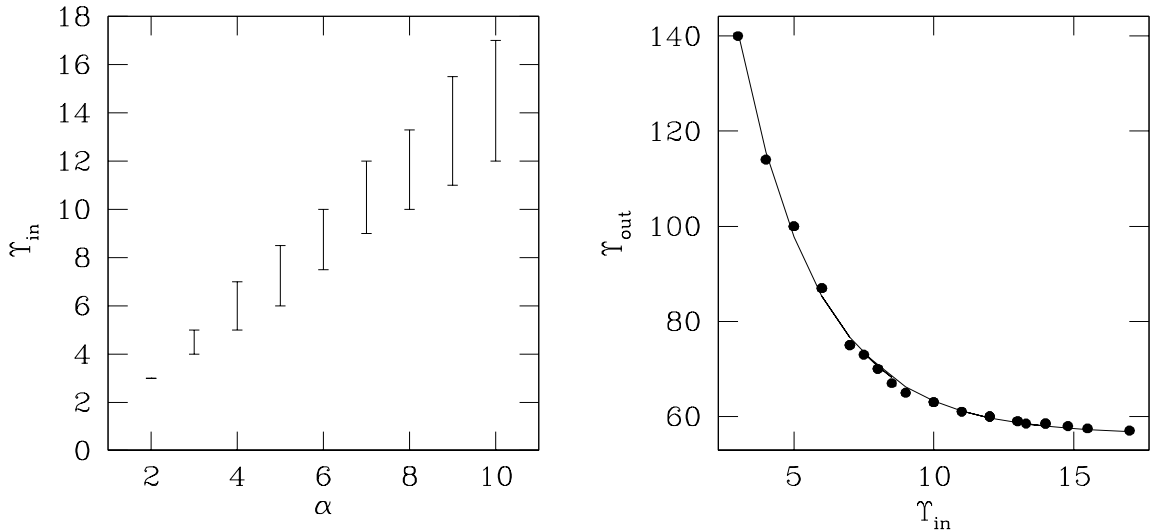,height=16.0truecm,width=16.5truecm}}
\vspace{-8cm}
\caption{\small{Left panel: the allowed ranges in $\Upsilon_{in}$, as
a function of the estimated central local $M/L$, as parametrised by
$\alpha$, for the Sersic model that we consider.  Top right panel:
$\Upsilon_{out}$ plotted as a function of $\Upsilon_{in}$; the solid
line represents an exponentially decaying function of $\Upsilon_{in}$
that is a very good fit to our experimentally obtained data.}}
\label{fig:1440}
\end{figure*}

For a given choice of the central $M/L$, a range of $\Upsilon_{in}$
are allowed. This is shown on the left panel of
Figure~\ref{fig:1440}. This range is evidently smaller for the Sersic
toy model, compared to the cored models. Again, we find that a
one-to-one relationship exists between $\Upsilon_{in}$ and
$\Upsilon_{out}$, as shown in the right panel of
Figure~\ref{fig:1440}.

Thus, it is apparent that galaxies with brightness distributions that
are best described by a Sersic profile, can also be tackled by the
proposed mass modelling scheme. The two main differences in the
formalism, as applicable to the cored and Sersic brightness profiles
are the following.
\begin{enumerate}
\item The luminosity density for a Sersic-type galaxy needs to be
normalised to 1000 L$_\odot$pc$^{-3}$, while a cored brightness profile
would require a normalisation factor that is 1 order of magnitude
smaller.
\item For the determination of $X_e$ of the Sersic-type galaxies, a
straight line fit to the plot of log$_{10}$(I) against $x^{1/4}$ needs
to be performed while for the cored galaxies, a linear fit is sought
to the relationship between the {\it core-removed} log$_{10}$(I)
profile and $x^{1/4}$.
\end{enumerate}

\section{Results}
\label{sec:results}
\begin{figure}
\centerline{
\psfig{figure=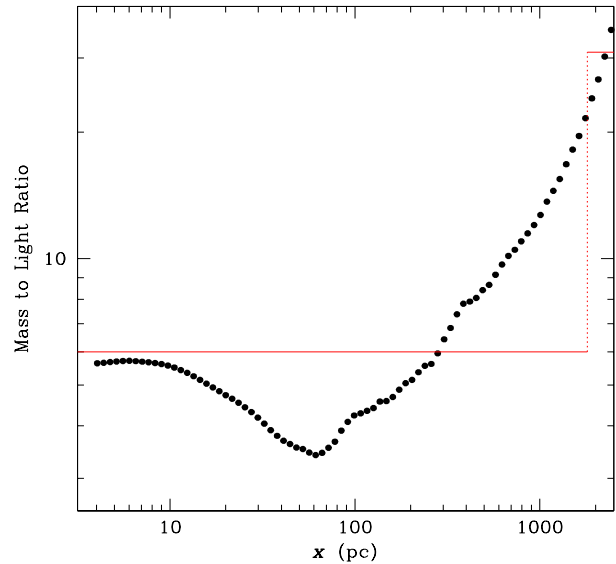,height=8.0truecm,width=8.5truecm}}
\caption{\small{Plot of the raw (in broken red lines) two-stepped
local $M/L$ ratio that is used to scale the luminosity density profile
of the toy galaxy of core radius 17.3pc, given model parameters of
$\alpha$=3, $\Upsilon_{in}$=6 and $\Upsilon_{out}$=31. The resulting
discontinuous mass density profile is then smoothed according
smoothing prescription mentioned above. When this smoothed mass
density distribution is compared to the deprojected luminosity density
profile, the resulting local $M/L$ profile is shown in filled
circles. As is apparent, the raw two-stepped $M/L$ distribution varies
significantly (beyond error bars) from the final smoothed $M/L$ ratio
distribution.}}
\label{fig:smooth_raw}
\end{figure}

\noindent
Figure~\ref{fig:smooth_raw} brings out the change in the local
mass-to-light ratio profile, as a result of the smoothing; the mass
density distribution that results from the smoothing of the scaled
luminosity density profile, is compared to the luminosity density, to
extract the smoothed local $M/L$. This is seen in the figure to be
different from the raw, two-step, discontinuous $M/L$ distribution
that is used to scale $L(x)$. The smoothing is noted to affect
mass-to-light on both sides of the jump radius.

\begin{figure*}
\centerline{
\psfig{figure=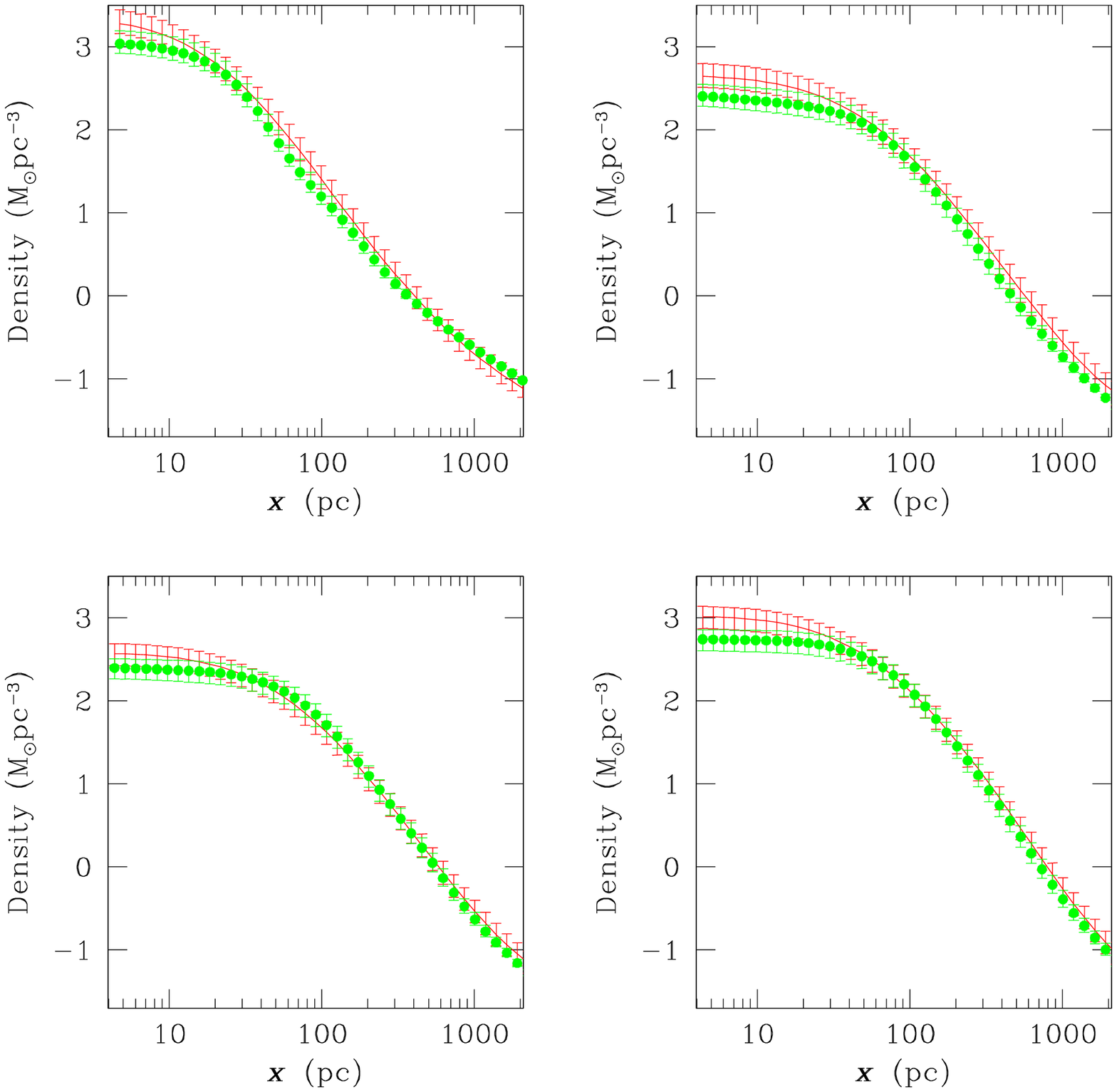,height=16.0truecm,width=16.5truecm}}
\caption{\small{Comparison of recovered (in red) and model (in green)
mass density profiles for the cored toy galaxies with varying $\alpha$
(lower panels) and core radii (upper panels). The two different core
radii corresponding to the plots on the upper right and left are about
17.3pc and 7pc, respectively. These toy configurations have been
ascribed the same $\alpha$ of 10. The lower right and left panels
correspond to the same $r_c$ of 20pc and $\alpha$=3 and 7,
respectively. The range of $x$ over which the profiles are shown
corresponds to thrice the $X_e$ for the $r_c$=20pc model. This is in
excess of 3$X_e$ for the models with the lower core radii. That the
overlap between the recovered and true profiles is notable over this
whole $x$ range, even in the cases of $r_c$=17.3pc and 7pc, reflects
the realisation that the mass modelling is often successful even
beyond $x=3X_e$. }}
\label{fig:varyrc}
\end{figure*}

Figure~\ref{fig:varyrc} displays the degree of overlap between the
estimated and true mass density profiles, for toy cored galaxies that are
normalised to the same central luminosity density and characterised by
varying shapes, i.e. varying $r_c$ (upper panels). The lower panels of
the same figure indicate a similar comparison, when $\alpha$ is made
to change. In all these cases, when we utilise the prescribed values
of $\Upsilon_{in}$ and $\Upsilon_{out}$, the predicted mass density
profile reproduces the true mass density profile very well, within
error bars.

\begin{figure*}
\centerline{
\psfig{figure=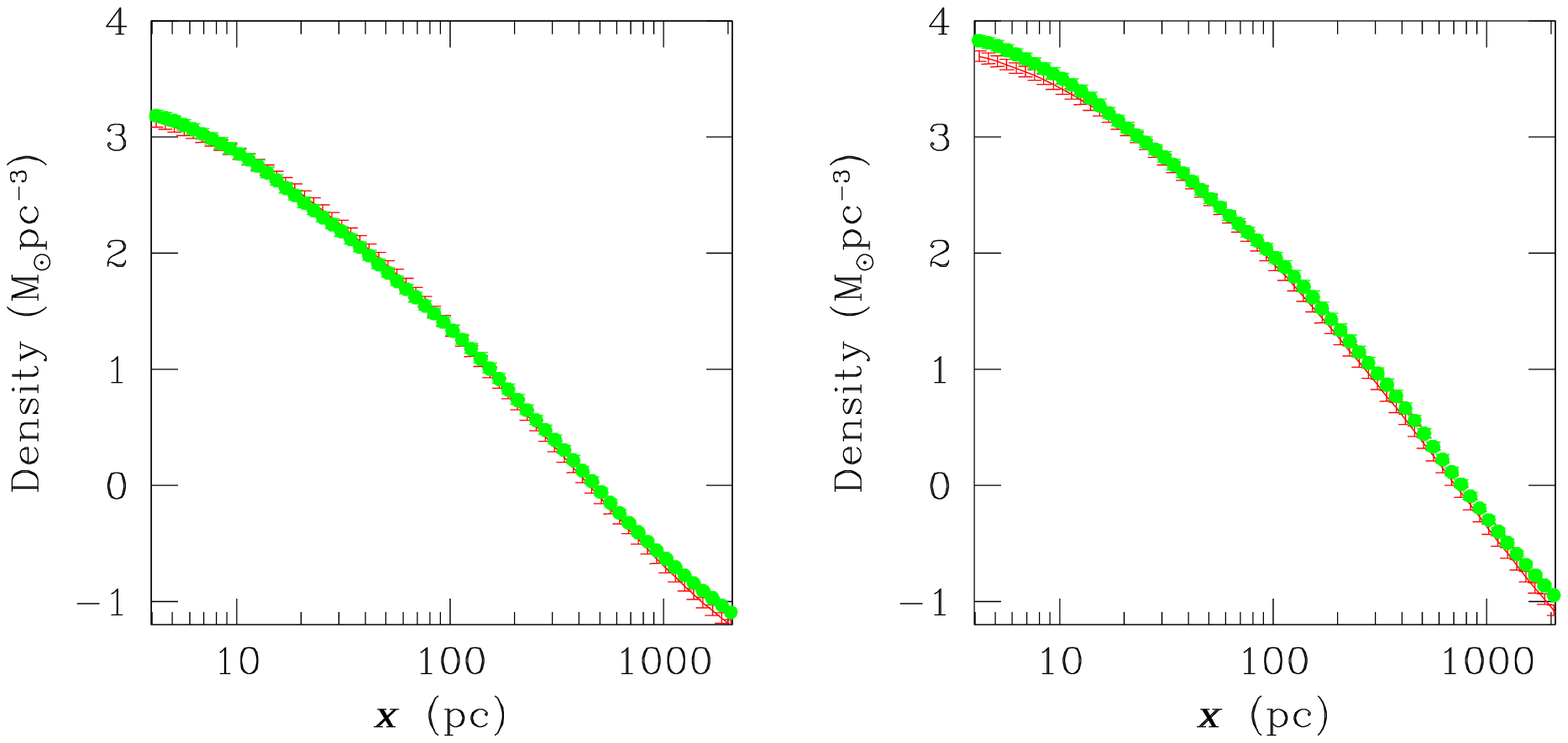,height=16.0truecm,width=16.5truecm}}
\vspace{-8cm}
\caption{\small{Similar to Figure~\ref{fig:varyrc}, except that in
this case, the comparison is shown for two distinct Sersic models that
have been ascribed $\alpha$ of 2 (left) and 9 (right). }}
\label{fig:sersic}
\end{figure*}

Figure~\ref{fig:sersic} brings out the resemblance between the known
mass density distribution along the major axis (in green) and its
predicted counterpart (in red), for the Sersic model investigated, for
a value of $\Upsilon_{in}$ that is picked from the range allowed for a
given choice of the estimated central $M/L$.

\section{Allowed galaxies}
\label{sec:allowed}
\begin{figure}
\centerline{
\psfig{figure=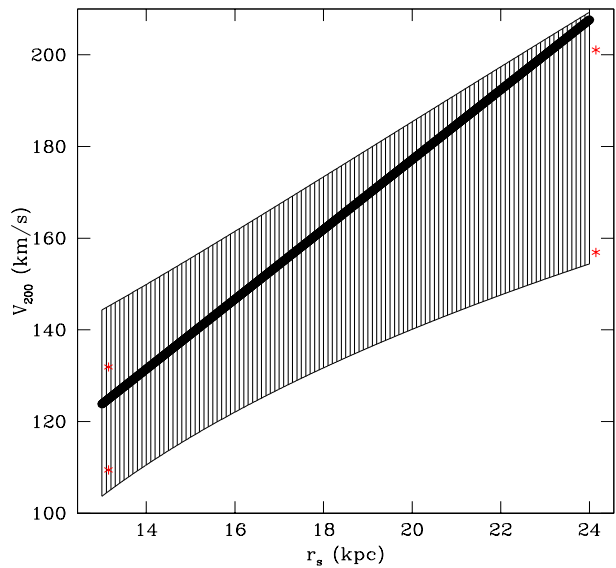,height=8.0truecm,width=8.5truecm}}
\caption{\small{The shaded region indicates the section of the
$V_{200}-r_s$ plane that corresponds to galaxy configurations that can
be successfully modelled with the proposed mass modelling scheme. The
solid black line is the NFW prediction, according to Hoekstra, Yee
$\&$ Gladders, (2004). The red stars mark the extremes of the ranges
in $V_{200}$ and $r_s$ that are allowed for the Sersic model
considered in the paper.}}
\label{fig:allowed}
\end{figure}

\noindent
This scheme of characterising the two-step local mass-to-light ratio
profile is found to work only as long as the galaxies at hand lie
within a certain band in the $M_s - r_s$ space, where $M_s$ is the
mass scale and $r_s$ is the scale radius that describe the NFW dark
halo mass density profile (see Equation~\ref{eqn:nfw}). This band is
translated into a range of the allowed values for the cored galaxies,
in the $V_{200} - r_s$ space, (where $V_{200}$ is the circular
velocity at radius $r_{200}$), in Figure~\ref{fig:allowed}.  The
broken line in this figure indicates the state of the NFW halos, as
given in \scite{hoekstra04}. The range in $M_s$ allowed for
Sersic-type galaxies is slightly more constricted, as compared to the
cored galaxies. 

To clarify, by these ``allowed ranges'', we imply the ranges in
$V_{200}$ and $r_s$, over which the same $\Upsilon_{in}$ and
$\Upsilon_{out}$ values are valid.

Thus, it is apparent that the mass modelling scheme that is being
proposed here is applicable even to systems that may exhibit a
significant range in concentration. At the same time, we note that
dwarf galaxies which typically have $V_{200}\sim100$ kms$^{-1}$ cannot
be characterised by this formalism.

In addition to these constraints on the $M_s$ and $r_s$, the galaxies
also need to be ``ordinary'' in the sense that they should,
qualitatively speaking, have a lower inner mass-to-light ratio than a
higher $M/L$ at higher radii. However, if the galaxy at hand is known
to have a central mass condensation from independent measurements
(such as M87), the raw two-step $M/L$ profiles discussed above
(described by a low $\Upsilon_{in}$ and a $\Upsilon_{out}$) will not
be apt. We refer to such galaxies as ``not ordinary'' in this
paper. For such a system, an alternative three-step, local
mass-to-light profile can be conjured. In such a configuration, the
value of the $M/L$ in the innermost region must be retained as high,
while the intermediate range is envisaged to be described by a low
$M/L$. Outside of the second jump radius, the dark matter contribution
is expected to kick in; therefore, the mass-to-light ratio in the
outermost part should be held at a high value. The values of the $M/L$
in the different parts of such systems are not expected to be given
according to the prescriptions mentioned above; nor are the two
relevant jump radii in these cases, expected to be given as multiples
of $X_e$ in the same way as for ``ordinary'' galaxies. 

In general, the mass modelling of such systems are beyond the scope of
the proposed formalism. Nevertheless, in Section~\ref{sec:m87}, we
will show how the basic concept behind our proposed mass modelling
formalism can be extended to model M87, using whatever we know of the
physics of M87, from literature.

\section{M87}
\label{sec:m87}
In this section, we present a mass model of the galaxy M87. We compare
our results with the enclosed mass profile put forward by
\scite{cote87} and the mass density of M87, suggested by
\scite{romkoch87}.

M87 (NGC 4486) is a cD galaxy that lies near the centre of the Virgo
cluster. It has been extensively studied both for the photometry of
its stars and its globular cluster system (\cite{strom81}) as well as
for their kinematics; some of the latest investigations are due to
\scite{romkoch87}, \scite{cote87}. M87 is marked by the existence of a
central supermassive black hole (\cite{machetto}, \cite{vdm94}) and a
dearth of dark matter in the outer parts; we configure the details of
the raw $M/L$ profile accordingly. 

Given that M87 is ``not ordinary'', according to the prescription
suggested in the previous section, it is expected to require a three
step discontinuous raw $M/L$ profile, in which the local $M/L$ in the
innermost step is high (in conjunction with the reported central
supermassive black hole) and that in the intermediate and outermost
steps are expected to be moderate and high respectively. However,
given that M87 has been modelled with no dark matter halo in the past
(\cite{mclaughlin99}), with most of its dark matter being that of
Virgo, it makes sense for us to maintain the same value of the local
$M/L$ in the outer two steps (at least upto the radius to which
photometric information is available). Thus, the raw $M/L$ profile is
rendered two stepped and we are required to judiciously choose the
jump radius $x_{in}$ that marks the end of the regime where the
influence of the central black hole dominates. The details of the
choice of this raw $M/L$ profile are discussed below:
\begin{itemize}
\item $x_{in}$ is chosen to be 0.2$^{''}$; the ACS surface brightness
profile of M87 indicates a steepening of the profile from about
0.2$^{''}$ inwards.
\item Outside the central 0.2$^{''}$, the mass to light ratio is
considered to be constant, with a moderately large value of about
10. This is maintained till the end of the radial range to which the
photometric data extends (about 100$^{''}$), which is just slightly
higher than the $X_e$ of about 96$^{''}$. We find that the allowed
range of this intermediate $M/L$ is 6 to 13.
\item The observed central line-of-sight velocity dispersion of M87
(\cite{vdm94}) is about 400kms$^{-1}$. This translates to an enclosed
mass of about 1.6$\times$10$^9$M$_\odot$. Comparison with photometry
implies a local central $M/L$ of about 200.
\end{itemize}

Now, let us discuss the predicted mass models for M87, against which
the recovered profile is checked. \scite{romkoch87} adopt an orbit
modelling approach to infer the mass density, from measurements of
velocities of stars (to about 1.5 effective radii) and globular
clusters (to about 5 effective radii).  The available surface
brightness and kinematic observations of the stars and globular
clusters over assorted radial ranges were used in this work.
\scite{romkoch87} use the Schwarzschild orbit library method to work
with generalised mass models that include a contribution from the
luminous matter; this is represented by the luminosity density
$\nu_{\star}(r)$ that is scaled by the $M/L$. The dark matter density
is considered to be NFW type. The fitting is conducted on the joint
constraints from the stellar and globular cluster velocity data, in a
Jeans equation formalism. In this work, the availability of only one
component of the velocity vector (the line-of-sight component), must
render a wide range of solutions possible.

\begin{figure*}
\centerline{
\psfig{figure=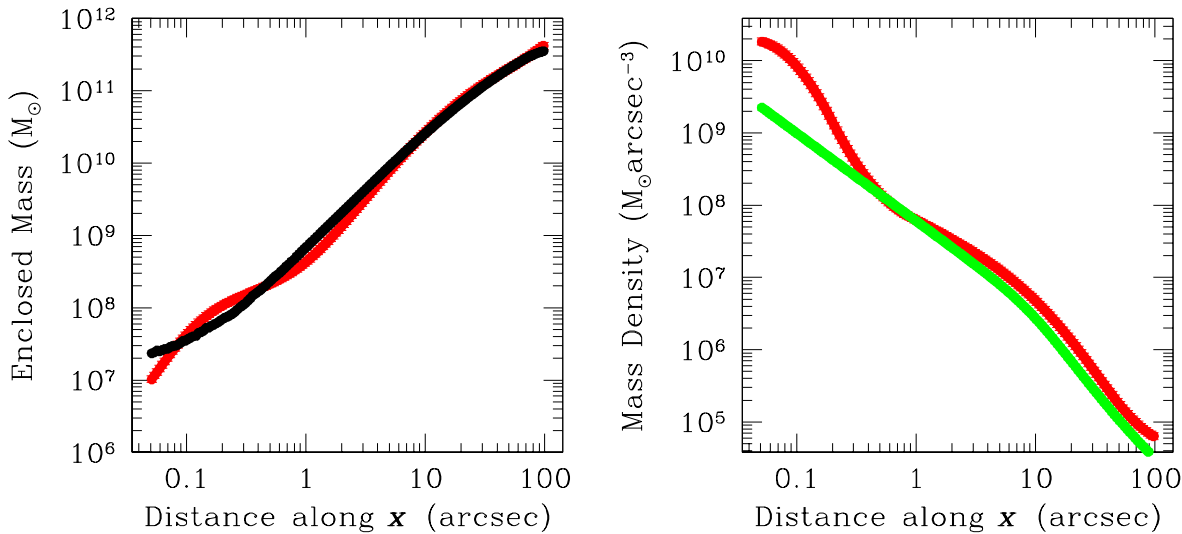,height=16.0truecm,width=16.5truecm}}
\vspace{-8cm}
\caption{\small{Figure showing the enclosed mass profile (left), and
mass density (right) of M87. The ACS photometry for M87 is
deprojected by DOPING to result in the luminosity density distribution
of this galaxy. The recovered luminosity density is then scaled by a
raw two-step local $M/L$ profile that has a jump radius of 0.2$^{''}$,
$\Upsilon_{in}$=200 and $\Upsilon_{out}$=6. The resulting
discontinuous mass density distribution is smoothed (two successive
applications of a boxcar smoothing filter of filter size 26 which
corresponds to the jump radius of 0.2$^{''}$). This smoothed mass
density distribution (in red) is compared (in the right panel) to that
predicted by Romanowsky $\&$ Kochanek (2001) (in green). The enclosed
mass profile that is calculated from this recovered mass density
profile is shown in red in the left and is compared to the predictions
of Cote et. al (2001) (in black).}}
\label{fig:m87_new}
\end{figure*}

The best fit model from \scite{romkoch87} is presented in the 
right panel in Figure~\ref{fig:m87_new} in green. Even this best
fit model is reported to suffer from the ailment that the values of
the normalisation of the M87/Virgo mass profiles, as obtained from the
stellar and globular data sets separately, are unequal. This could be
explained by either a larger, less concentrated halo or by the
relaxation of the assumption of sphericity. In fact, the ACSVCS image
of M87 shows clearly that the eccentricity in M87 is not zero; the
eccentricity is seen to dip sharply in the inner regions and then
appears to tend towards a maximum of about 0.4, as radius increases.

The dynamical analysis of the radial velocities of 278 globular
clusters of M87 is presented in \scite{cote87}. The total (projected)
radial range spanned by the kinematic data set of \scite{cote87}
extends from about 36$^{''}$ to about 6.5 effective radii. This
kinematic data is then fed into the Jeans equation under the
assumption of a {\it spherical geometry}. In order to evade the
problem of mass-anisotropy degeneracy, \scite{cote87} take the
distribution of the total mass of M87 and the Virgo cluster that was
deduced by \scite{mclaughlin99}. This enclosed mass distribution is
shown in black in the left panel of
Figure~\ref{fig:m87_new}. In this model, the dark matter is assigned
wholly to the cluster. 

The surface brightness and eccentricity profile of M87, observed as
part of the ACS Virgo Cluster Survey were used as inputs to DOPING
which recovered a luminosity profile for the galaxy. When the inferred
light distribution is scaled by the two step $M/L$ profile discussed
above, and the consequent discontinuous mass density distribution is
smoothed (according to the smoothing prescription suggested in
Section~\ref{sec:smooth}; smoothing filter size corresponds to $x_{in}$).
The smoothed mass density profile is used to extract the enclosed mass
profile, which is compared to the enclosed mass distribution of M87
that has been predicted by \scite{cote87} (see
Figure~\ref{fig:m87_new}). The comparison appears favourable indeed
though the comparison of the recovered mass density to that predicted
by \scite{romkoch87} is less encouraging in the central parts of the
galaxy. This discrepancy however does not reflect badly on our mass
modelling technique since the density predicted by \scite{romkoch87}
is also in deficit of the enclosed mass predictions of \scite{cote87},
due to the lack of inclusion of a central black hole in their model.

\section{Summary and Conclusions}
\label{sec:discussions}
\noindent
We propose a scheme to retrieve the total mass density to about 3
times the effective radius in ellipticals, using nothing but photometry
and the central velocity dispersion: the exact two measurements that
are most commonly available for most galaxies. The basic framework of
this formalism relies upon the construction of a two-step,
discontinuous, local $M/L$ profile, which is chosen to suffer its
discontinuity at 3$X_e$. Here, $X_e$ is the major-axis coordinate
equivalent of the effective radius, as defined in terms of the fitting
of the observed brightness profile to the de'Vaucouleurs $R^{1/4}$
law. Prescriptions are offered for constraining the inner and outer
amplitudes of this sought two-step $M/L$ profile. While
$\Upsilon_{in}$ is found to lie within an allowed range of
mass-to-light values for a given estimate of the central mass-to-light
in the system, $\Upsilon_{out}$ is fixed for a given $\Upsilon_{in}$,
for a galaxy of a given shape. The details of the prescriptions are
found to differ slightly, depending on whether the observed surface
brightness is noted to be cored or Sersic in nature.

This formalism is designed to accommodate for the usual problems that
are associated with our lack of knowledge of the state of anisotropy
in the system and deviations from sphericity. Thus, the measured
central velocity dispersion is implemented in the virial theorem to
pin down the (local) central mass-to-light ratio, but the possibility
of miscalculating the same due to anisotropy always exists. However,
this mistake can be covered for, to some extent, in the sense that as
long as the true central mass-to-light ratio ($\Upsilon_{in}$) lies
within a range above the estimated central $M/L$ (parametrised by
$\alpha$), the recovered mass density profile is found to be
compatible with the true distribution. This range is given as
2$\alpha$, the central luminosity density of which is normalised to
100 L$_\odot$pc$^{-3}$. If on the other hand, the brightness profile
is Sersic type, the normalisation factor is 1000 L$_\odot$pc$^{-3}$
and for the only Sersic model that was explored, the upper bound on
the allowed range for $\Upsilon_{in}$ is relatively more constricted,
viz. 1.65 times $\alpha$. A one-to-one relationship between
$\Upsilon_{in}$ and $\Upsilon_{out}$ exists in both cases.

Such a prescription can fail if the degree of anisotropy in the centre
of the system is so high that the virial estimate of the central mass
density is rendered different from the true central density by a
factor in excess of 2$\alpha$ in the case of cored galaxies and
1.65$\alpha$, for Sersic galaxies. The good news is that when this
happens, the formalism can indicate the same; such configurations
cannot be handled by the proposed scheme. However, it is unlikely that
the galaxy will be this anisotropic at the centre since a careful
analysis of a sample of over 2000 galaxies in SDSS indicates that the
dynamical mass enclosed with 1$R_e$ is about 1.82 times the virial
mass estimate, as given in Equation~\ref{eqn:virial} (using the
result of \cite{padmanabhan04}).

An important factor that dictates the success of our mass modelling
scheme is smoothness of dispersion profiles in real galaxies. As
\scite{gebhardt00} suggested, the largest variations in normalised
dispersions occur in galaxies, either in the inner parts ($R <
0.5X_e$) or at radii exceeding 2$X_e$. In the radial range of $0.5X_e
< R <2X_e$, the variations in dispersion profiles is reported to be
less than 5$\%$. \scite{gerhardkronawitter} suggest that the $M/L$
profile starts to pick up around 0.5-2$X_e$. This uniform dynamical
characteristic of real ellipticals is exploited in the production of
the formalism.

\subsection{Core Fitting Method}
\noindent
At this point, it merits mention that though one part of the proposed
mass modelling technique resembles King's core fitting method
(\cite{king66}, \cite{rood72}) in some way, the whole of the advanced
technique is very different from the core fitting method. The
resemblance lies in the invoking of the virial theorem in the
extraction of the mass that lies enclosed within a pre-fixed radius;
in our work, this is the minimum radius ($x_0$) at which a velocity
dispersion measure is available. In the implementation of the virial
theorem, isotropy and sphericity are assumed, as in the core fitting
method.

But the essential difference between the two methods lies in the fact
that while the virial estimate of the mass is advanced as the answer
in the core-fitting method, in our method, this mass is used to spot
the range of allowed $\Upsilon_{in}$ values that correspond to this
estimate. In other words, given a virial estimate of the mass within
$x_0$, we attempt to decipher the range of values within which the
true central (local) $M/L$ lies. The availability of a range in this
$M/L$ offers some respite from the difficulties caused by the possible
existence of anisotropy, which the core-fitting method is not
safeguarded against; in fact, it appears from work done with real
galaxies that this respite is realistic indeed. Moreover, our method
would forewarn us in case the system is so severely anisotropic that
the method would fail.  The core-fitting method has no such inbuilt
safeguard.

Most importantly, once we have chosen our $\Upsilon_{in}$, (which then
fixes $\Upsilon_{out}$), smoothing is implemented. The smoothed mass
density distribution is then found to correspond to a local $M/L$
profile that is found to be {\it different} from the choice of
$\Upsilon_{in}$, in the inner parts of the galaxy. The whole of the
$M/L$ distribution, from $x=0$ to $x=3X_e$, is rendered affected by
the overall smoothing. This is clearly borne by the
Figure~\ref{fig:smooth_raw}.  On the other hand, the $M/L$ in the
central parts of the galaxy, as found from the core-fitting method,
would of course be just what the virial estimate indicates.

All in all, we can say that our method does invoke the virial theorem
to translate velocity dispersion information into an enclosed mass
value, as is done in the core-fitting method, but in our technique, we
then go ahead to refine that virial estimate in multiple ways, unlike
in the core-fitting method. The proposed mass modelling formalism is
applicable to aspherical galaxies that are at least as anisotropic in
the centre as massive ellipticals in the SDSS, even when these systems
bear constant surface brightness cores, and it provides a mass
distribution over a radial range extending to about 3 times the
effective radius. The core-fitting method is much more constrained in
the sense that it provides the mass profile only in the central
regions of galaxies that do not harbour a constant surface brightness
core and which when aspherical and/or anisotropic, cause the estimated
mass distribution to be erroneous.

\subsection{The Case of M87}
\noindent
The fundamental motivation for the raw two-stepped local $M/L$ profile
to be ascribed a low inner and high outer $M/L$ amplitude is to
replicate the anticipated trend of the dark matter kicking in at large
radii (3$X_e$ and higher). But if a system is known (from independent
measurements) to harbour a central mass condensation then the
innermost radial range should be assigned a high local $M/L$
indeed. Such is the case for the galaxy M87. Additionally, M87 is
known to be a cD galaxy lying at the centre of Virgo. Thus, it can be
modelled as being bereft of any dark matter, with all the dark matter
ascribed to Virgo itself (\cite{mclaughlin99}). Such information
about the reported state of the matter distribution in M87 can
motivate us to model this galaxy, using our mass modelling scheme,
with a two step $M/L$ profile in which the inner $M/L$ amplitude is
high, compared to the rest of the radial range covered by
photometry. Once we configure our raw $M/L$ profile in this way, the
recovered mass distribution in M87 is found to closely follow the mass
models predicted for M87 from earlier studies. Thus, the proposed
technique relies upon the inclusion of as much measured information
about the galaxy at hand, as is possible.

\subsection{Connection between luminous $\&$ dark matter densities}
\noindent 
Once $\Upsilon_{in}$ is constrained, $\Upsilon_{out}$ is fixed; this
correlation between $\Upsilon_{in}$ and $\Upsilon_{out}$ depends on
whether the system is cored or Sersic. A full investigation was done
with the cored galaxies that were normalised to a central luminosity
density of 100 L$_\odot$pc$^{-3}$, to suggest that
\begin{eqnarray}
\Upsilon_{out} &=& \displaystyle{
                               8.32\times10^5r_c^{-3.75}} +\\ \nonumber
            {}   && \displaystyle{9.76\times10^8r_c^{-5.31}
                               \exp\left[\frac{-\Upsilon_{in}}{1.005^{-6}r_c^4+1.85}\right]}. 
\end{eqnarray}
Similar analysis performed with a toy Sersic brightness profile
revealed that $\Upsilon_{out}$ is again related to $\Upsilon_{in}$ via
an exponential decay though a more thorough investigation needs to be
undertaken to appreciate the dependence of this relation on the Sersic
parameters.

\begin{figure}
\centerline{
\psfig{figure=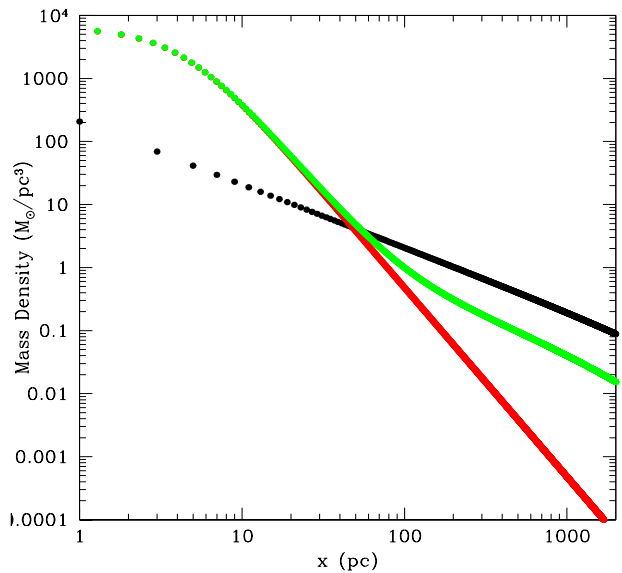,height=8.0truecm,width=8.5truecm}}
\caption{\small{Figure showing trends in dark matter density (black)
and luminous matter density (in red) for a model cored galaxy with a
core radius of about 14.4pc, eccentricity 0.8 and $\alpha$=10. The sum
of these two components is shown in green.  The $x$ range spanned in
the plot corresponds to thrice the value of $X_e$ for this model
galaxy.  }}
\label{fig:analytical}
\end{figure} 

The fact that the local mass at 3$X_e$ in our toy ellipticals is
related uniquely to the mass near the centre, betrays a relationship
between the dark matter and luminous matter content in these systems.
Given that the total mass density in these models has a contribution
from an NFW-type dark halo and $\alpha$ times a (cored or Sersic)
luminosity density profile, (where $\alpha$ is a scalar), such a
relationship is not altogether unexpected. As shown in
Figure~\ref{fig:analytical}, it is clear that in the inner regions
($x\leq{r_c}$) the total mass density is dominated by the shape of the
underlying luminosity density profile, while the dark halo
predominates in the outer regions ($x\sim 3X_e$). Thus, when the
luminous mass is low, the dark mass is high and the vice versa. This
trend manifests itself into the exponential decay of the outer $M/L$
value as the inner $M/L$ is increased. It must be kept in mind that
this relationship between $\Upsilon_{in}$ and $\Upsilon_{out}$ is in
the context of the $raw$ mass-to-light distribution, while the
Figure~\ref{fig:analytical} refers to the situation in the smoothed
configuration. 

Such a fall-off is also expected for the Sersic galaxies, though the
exact nature of the trend with an intrinsic scale length (such as the
core radius in case of the cored galaxies) would expectedly differ.
But of course, that our {\it recovered} profiles exhibit this
exponential fall-off, indicates the success of our modelling. 

\subsection{Limitations $\&$ Outreach}
\noindent
The more pertinent question is, how general is this trend between the
local $M/L$ at the centre and at 3$X_e$. In other words, how
convincing is the idea that in general there is a quantifiable
connection between the luminous matter density in elliptical galaxies
and the dark matter content at approximately 3 times the effective
radius?

The answer obviously depends on the genericness of the models that we
have consulted in our work. As far as the amplitude of the model
luminosity densities are concerned, it has been constrained to a value
that is typical of the order of magnitude of the central density of
real ellipticals. Also, a range of core radii has been scanned
successfully in our cored models. Though a range of Sersic type model
brightness profiles have not been scanned, the trends noted with the
single model Sersic galaxy that was considered here, appear similar in
nature to those found for the cored models. Thus, it seems possible
that the inner and outer $M/L$ in the raw two-stepped $M/L$ profile
will be connected even in the Sersic case, with the exact nature of
this connection given by parameters intrinsic to the brightness
distribution.

All this corroborates the generality of our models. At the same time
it needs to be appreciated that the success of the procedure depends
on the conformity of a real elliptical galaxy to the toy models that
were used to set up the rules that connect the inner and outer
mass-to-light ratios. In particular, the surface brightness
distribution of the real galaxy and the compatibility of its dark halo
to the NFW form need to be checked. In fact for each unique family of
brightness profiles, a different rule connecting $\Upsilon_{in}$ and
$\Upsilon_{out}$ is expected to exist. In this paper, we presented the
same for cored galaxies that were normalised to the same central
luminosity density and found a similar trend with one toy Sersic
galaxy. An extensive study of toy Sersic galaxies is underway, to form
a rule similar to the one prescribed in Equation~7. This proposed
project will also include an investigation of the effect of changing
the nature of the dark matter distribution in the model galaxies, on
this rule. Thus, this method can potentially provide a simple
automated way of deciphering mass profiles of a large sample of
elliptical galaxies.

Thus one important conclusion of this paper is that a connection
exists between the inner luminous matter and the dark matter density
at about 3$R_e$ in cored ellipticals and perhaps also in Sersic
galaxies. This notion, though encouraged by intuition, needs to be
cultivated better in order to ascertain the physical origin of the
accompanying parametrisation.

Similar relationships have been explored in the context of disk
galaxies; some of more recent works include contributions from
\scite{sancisi04}, \scite{verheijen01} and \scite{noordermeer06}. In
these studies and the many more that preceeded them, a relationship
was noted between the characteristics of the rotation curves and the
optical luminosities of disk galaxies.  Investigations along similar
lines are of course not possible with ellipticals but studies of
samples of ellipticals in various catalogues have been undertaken to
understand mass distribution in early type systems (\cite{mamon05},
\cite{padmanabhan04}, \cite{lintott06}, among others). The mass
modelling technique that is advanced in this paper is one way to
pursue the same question, over larger radial ranges than are typically
covered in these studies (about 3 effective radii as compared to 1).

Given the simplicity of this method and the twosome observational
input required, (which are readily provided by most catalogues,
including the SDSS), a future project (\cite{dolfdal}) is planned in
which the construction of the mass build up in the nearby universe
will be attempted.

\section*{Acknowledgements}
{The author is indebted to Michael Merrifield, Chris Conselice and Dolf
Michielsen for their helpful comments, and to the anonymous referee
whose criticisms and comments helped to shape the work better. The
author is funded by a Royal Society Fellowship.}

\vspace{1cm}
\bibliographystyle{mnras}
\bibliography {dc2}
\end{document}

@article{zepf00,
	author=``Zepf, S. E. and Beasley, M. A. and Bridges, T. J. and  Hanes, D. A.and  Sharples, R. M. and  Ashman, K. M. and Geisler, D.",
	title="",
	journal=Astronomical Journal,
	year=2000,
	volume=120,
	pages=2928	}

@article{king66,
	author="King, I. R.",
	title="",
	journal=aj,
	year=1966,
	volume=71,
	pages=64	}

@article{rood72,
	author="Rood, H. J. and Page, T. L. and Kintner, E. C. and King, I. R",
	title="",
	journal=apj,
	year=1972,
	volume=179,
	pages=627	}

@article{hoekstra04,
	author="Hoekstra, H. and Yee, H. K. C. and Gladders, M. D. ",
	title="",
	journal=apj,
	year=2004,
	volume=606,
	pages=67	}

@article{lintott06,
	author="Lintott, C. J. and Ferreras, I. and Lahav, O.",
	title="",
	journal=apj,
	year=2006,
	volume=648,
	pages=826	}

@article{padmanabhan04,
	author="Padmanabhan, N. and Seljak, U. and Strauss, M. A. and Blanton, M. R. and Kauffmann, G. and Schlegel, D. J. and Tremonti, C. and Bahcall, N. A.and Bernardi, M. and Brinkmann, J. and Fukugita, M. and Ivezić, Ž.",
	title="",
	journal=newast,
	year=2004,
	volume=9,
	pages=329	}

@article{verheijen01,
	author="Verheijen, M. A. W.",
	title="",
	journal=apj,
	year=2001,
	volume=563,
	pages=694	}

@article{noordermeer06,
	author="Noordermer, E.",
	title="",
	journal=apj,
	year=2006,
	volume="",
	pages="in preparation"	}

@article{buote,
	author="Buote, D. A. and Jeltema, T. E. and Canizares, C. R. and Garmire, G. P. ",
	title="",
	journal=apj, 
	year=2002,
	volume=577,
	pages=183	}

@article{vdm94,
	author="van der Marel, R. P.",
	title="",
	journal=mnras, 
	year=1994,
	volume=270,
	pages=271	}

@inproceedings{sancisi04,
        author="Sancisi, R.",
        title="",
        booktitle="International Astronomical Union Symposium no. 220",
        editor="Ryder, S. D. and Pisano, D. J. and Walker, M. A. and Freeman, K.",
        year="2004",
        publisher="ASP, San Francisco",
        pages="233",
        notes=""        }

@article{bertola,
	author="Bertola, F. and Pizzella, A. and Persic, M. and Salucci, P.",
	title="",
	journal=apj, 
	year=1993,
	volume=416,
	pages="45L"	}

@article{gebhardt00,
	author="Gebhardt, K. and Bender, R. and Bower, G. and Dressler, A. and Faber, S. M. and Filippenko, A. V. and Green, R. and Grillmair, C. and Ho, L. C. and Kormendy, J. and 5 coauthors",
	title="",
	journal=apj, 
	year=2000,
	volume=539,
	pages="13L"	}

@article{strom81,
        author="Strom, S. E. and Strom, K. M. and Wells, D. C. and Forte, J. C. and Smith, M. G. and Harris, W. E.",
        title="",
        journal=apj,
        year=1981,
        volume=245,
        pages=416       }

@article{dolfdal,
	author="Michielsen, D. and Chakrabarty, D.",
	title="",
	journal=mnras, 
	year=2006,
	volume="",
	pages="in preparation",
        notes=""	}

@article{machetto,
	author="Macchetto, F. and Marconi, A. and Axon, D. J. and Capetti, A. and Sparks, W. and Crane, P.",
	title="",
	journal=apj, 
	year=1997,
	volume=489,
	pages=579	}

@article{napolitano05,
	author="Napolitano, N. R. and Capaccioli, M. and Romanowsky, A. J. and Douglas, N. G. and Merrifield, M. R. and Kuijken, K. and Arnaboldi, M. and Gerhard, O. and Freeman, K. C",
	title="",
	journal=mnras, 
	year=2005,
	volume=357,
	pages=691	}

@book{bible,
      author="Binney, J. and Tremaine, S.",
	title="Galactic Dynamics",
	year=1987,
	publisher="Princeton University Press Princeton New Jersey"	}

@article{gerhardgenzel,
	author="Genzel, R. and Pichon, C. and Eckart, A. and Gerhard, O. E. and Ott, T.",
	title="",
	journal=mnras,
	year=2000,
	volume="317",
	pages="348",
        notes=""	}

@article{doping06,
	author="Chakrabarty, D. and Ferrarese, L.",
	title="",
	journal=aj,
	year="2006",
	volume="",
	pages="submitted to AJ",
        notes=""	}

@article{laura06,
	author="Ferrarese, L. and C\^ot\'e, P. and Jordan, A. and Peng, E. W. and Blakeslee, J. P. and Slawomir P. and Mei, S. and Merritt, D. and Milosavljević, M. and  Tonry, J. L. and West, M. J.",
	title="",
	journal=apjs,
	year=2006,
	volume="164",
	pages="334",
        notes=""	}

@article{cote03,
        author="C\^ot\'e, P. and McLaughlin, D. E. and Cohen, J. G. and Blakeslee, J. P.",
        title="",
        journal=apj,
        year=2003,
        volume=591,
        pages=850}

@article{cote87,
        author="C\^ot\'e, P. and McLaughlin, D.E. and Hanes, D.A. and Bridges, T.J. and Geisl\
er, D. and Merritt, D. and Hesser, J.E. and Harris, G.L.H. and Lee, M.G.",
        title="",
        journal=apj,
        year=2001,
        volume=559,
        pages="828"}

@article{mclaughlin99,
        author="McLaughlin, D.E.",
        title="",
        journal=apj,
        year=1999,
        volume=512,
        pages="L9"}

@article{romkoch87,
        author="Romanowsky, A. J. and Kochanek, C. S.",
        title="",
        journal=apj,
        year=2001,
        volume=553,
        pages="722"}

@article{mamon05,
        author="Mamon, G. A. and {\L}okas, E. L.",
        title="",
        journal=mnras,
        year=2005,
        volume=362,
        pages="95"}

@article{pnsscience,
        author="Romanowsky, A. J. and Douglas, D. N. and Arnaboldi, M. and Kuijken, K. and Merrifield, M. R. and Napolitano, N. R. and Capaccioli, M. and Freeman, K. C.",
        title="",
        journal="Science",
        year=2003,
        volume=301,
        pages="1696"}

@inproceedings{romanowsky04,
        author="Romanowsky, A. J. and Douglas N. G. and Kuijken, K. and Merrifield, M. R.
 and Arnaboldi, A. and Napolitano, N. R. and Merrett, H and Capaccioli, M.",
        title="",
        booktitle="Dark Matter in Galaxies, Proc. IAU Sumposium 220",
        editor="Ryder, S. and Pisano, D. J. and Walker, M. and Freeman, K.",
        year="2004",
        publisher="ASP, San Francisco",
        pages="165",
        notes=""        }

@article{gerhardkronawitter,
        author="Gerhard, O. and Kronawitter, A. and Saglia, R. P. and Bender, R.",
        title="",
        journal=aj,
        year=2001,
        volume="121",
        pages="1936"    }

@article{dalpras,
        author="Chakrabarty, Dalia and Saha, Prasenjit",
        title="",
        journal=aj,
        year=2001,
        volume="122",
        pages="232"     }

@article{dalzwart,
        author = "Chakrabarty, D. and Portegies Zwart, S.",
        title = "",
        journal = aj,
        year = 2004,
        volume = 128,
        pages = 1046 }

@article{merritt93,
        author="Merritt, D.",
        title="",
        journal=apj,
        year=1993,
        volume="413",
        pages="79"}

@article{merritt96,
        author="Merritt, D.",
        title="",
        journal=aj,
        year=1996,
        volume="112",
        pages="1085"}

\end{document}

%% file: abstract.tex
\noindent
Mass modelling of early-type systems is a thorny issue; even for the
few close by galaxies for which kinematic data is available, the
implementation of this data can get embroiled in problems that are
hard to overcome, unless more complete data sets are available. The
mass-anisotropy degeneracy is a typical example of this. In this
paper, we present a new mass modelling formalism for ellipticals that
invokes no other observations other than the central velocity
dispersion ($\sigma_0$) and photometry. The essence of the method lies
in choosing a local mass-to-light ratio ($M/L$) profile for a galaxy,
with which the deprojected luminosity density distribution (along the
major axis coordinate $x$) is scaled. The resulting discontinuous mass
density profile is then smoothed, according to a laid out
prescription; the local $M/L$ profile that stems from this smoothed
mass density, is found to be significantly different from the raw
$M/L$ distribution. A suite of model galaxies (both Sersic and cored
in nature) is used for intensive experimentation in order to
characterise this raw $M/L$ profile and in each case, the mass density
recovered from this mass modelling technique is compared to the known
mass distribution.  We opt to work with a raw $M/L$ profile that is a
simple two-stepped function of $x$, with a low inner and higher outer
value of $M/L-\Upsilon_{in}$ and $\Upsilon_{out}$, respectively. The
only constraint that we have on this profile is in the centre of the
galaxy, via $\sigma_0$. This value of $\sigma_0$ is implemented in the
virial theorem to obtain an estimate of the central $M/L$ ratio of the
galaxy. The fallibility of the virial mass estimate is taken care of,
by allowing for a range in the values of $\Upsilon_{in}$ that can be
used for a given galaxy model. Moreover, our experiments indicate that
$\Upsilon_{out}$ is uniquely known, for a given $\Upsilon_{in}$; for
cored galaxies, this functional form is found uniquely dependant on
the core radius. The physical basis for such a connection to exist
between the inner and out $M/L$ is discussed. The jump radius of the
raw $M/L$ profile is chosen to be about thrice the effective radius of
the galaxy. In this way, the local $M/L$ distribution is completely
specified and mass profiles of ellipticals can be constructed till
about 3 times the effective radius. The proposed technique is extended
to predict a mass profile for M87 which is then successfully compared
to distributions reported earlier from kinematic analyses.